\documentclass[aps,prb,manuscript]{revtex4}
\usepackage{epsfig}
\begin{document}
\title{Exact Stripe, Checkerboard and Droplet Ground States in Two Dimensions}
\author{
Zsolt~Gul\'acsi$^{a}$ and Miklos~Gul\'acsi$^{b}$}
\affiliation{
$^{(a)}$ Department of Theoretical Physics, University of Debrecen,
H-4010 Debrecen, Hungary\\
$^{(b)}$ Department of Theoretical Physics, Institute of Advanced Studies,
Australian National University, ACT 0200 Canberra, Australia }
\date{\today}
\begin{abstract}
Exact static non-degenerate stripe and checkerboard ground states
are obtained in a two dimensional generalized periodic Anderson model, 
for a broad concentration range below quarter filling.
The random droplet states also present in the degenerate ground state, are
eliminated by extending the Hamiltonian with terms of different
physical origin such as dimerization, periodic charge displacements,
density waves or distortion lines.
\end{abstract}
\maketitle

\section{Introduction}

Quasi-one-dimensional intrinsic inhomogeneities called stripes \cite{stucond}
represent of the most challenging problems in understanding self-organized
structures ever since the discovery of high $T_c$ cuprates \cite{stu3}.
Such textures have been observed in other than superconducting cuprates,
most notably in manganites \cite{stu4}, nickelates \cite{stu5} and
rare-earth compounds \cite{stu6}. Most recently it has been shown
that all cuprates exhibit intrinsic inhomogeneities in some form
or another: checkerboard structures have been identified in lightly
hole-doped copper oxides \cite{nat3}, while the electron-doped
cuprates show droplet formations \cite{stu8,stu9}. These textures are
so predominant in high $T_c$s, that
it has been even suggested \cite{nat1,nat2} as the best
candidate for the electron pairing mechanism in these materials.

After the existence of stripes have been established
in a broad spectrum of materials, it becomes clear that they emerge
due to a microscopic incompatibility between two phases
\cite{stu3,nat3,nat1,nat2}. However, this interplay
not always leads to stripe formations, but could also lead
to droplets (or blobs) \cite{stu8,stu9} as well.
Macroscopic inhomogeneity cannot occur as long as long-range forces are
present. But when such forces are absent, depending on doping, local
inhomogeneities appear in form of stripes or other form of clustering, e.g.,
droplets \cite{stu9,stu10}.

Even thought stripes have been known to exist much before high $T_c$
has been discovered \cite{stu2}, their theoretical understanding lacks
rigorous description. Hereafter we want to fill this gap. Hence
we present an exact solution of a generalized periodic Anderson model (PAM)
with ground states exhibiting intrinsic inhomogeneities of the stripe,
checkerboard and droplets type. Using a two band model as
a starting point of our analyses it even renders the obtained results
to be more generally applicable, since real materials are mostly
of multi-band type. These models are usually addressed by projecting
the multi-band structure into a few-band picture \cite{stu2new}, which
we stop for mathematical convenience, at a two band  level. However,
our study is not a simple two-band model but rather an extended PAM
which contains the added feature of strong correlation
effects originating from the on-site Coulomb repulsion $U$ present in the
correlated ($f$) band, an arbitrary $U_d > 0$ extra Hubbard interaction in
the free ($d$) band, leaves our results unchanged.

Since stripes and checkerboards are observed in a broad spectrum of materials,
we are primarily focusing on ground states which exhibit these inhomogeneities,
being less interested in the properties of the homogeneous phases in
which they appear.

Our exact results can be summarized as follows. Below quarter filling, two
stripe ground states emerge. One ($I$) with insulating and
paramagnetic stripes, while the second one ($II$) with itinerant and
ferromagnetic stripe lines. This second solution allows checkerboard
structures as well. In both cases, i.e., $I$ and $II$, the
inter-stripe line regions contain empty sites, hence are insulating.
The obtained ground states are in general degenerate. The degeneracy
is provided by a random blob structure corresponding to the same
energy, the blobs (random shape clusters) possessing the same properties as the
stripe lines in cases $I$ and $II$. The degeneracy of the ground state
can however be lifted, the resulting non-degenerate ground state
remaining of pure stripe or checkerboard character.
The lifting factor we find may have different physical origin as
distortion lines, dimerization or periodic charge displacement
(density waves). The obtained stripe formation processes are generic and
are less sensitive to the properties of the homogenous phases present at
quarter filling. We further note that marginal to the stripe problem, but
interesting for the PAM itself, we were able to prove rigorously that the
studied 2D PAM is ferromagnetic at quarter filling in a restricted region
of its parameter space. Similar result has been recently reported for 3D PAM 
as well \cite{pam1}.

The remaining part of the paper is structured as follows. Sec.II presents
the exact transformation of the Hamiltonian in a positive semidefinite form,
Sec.III. describes the obtained ground state solutions, Sec.IV. presents a
discussion of the obtained results, and Sec.V. concluding the paper, closes the
presentation.

\section{The transformation of the Hamiltonian}

\begin{figure}[h]
\centerline{\epsfbox{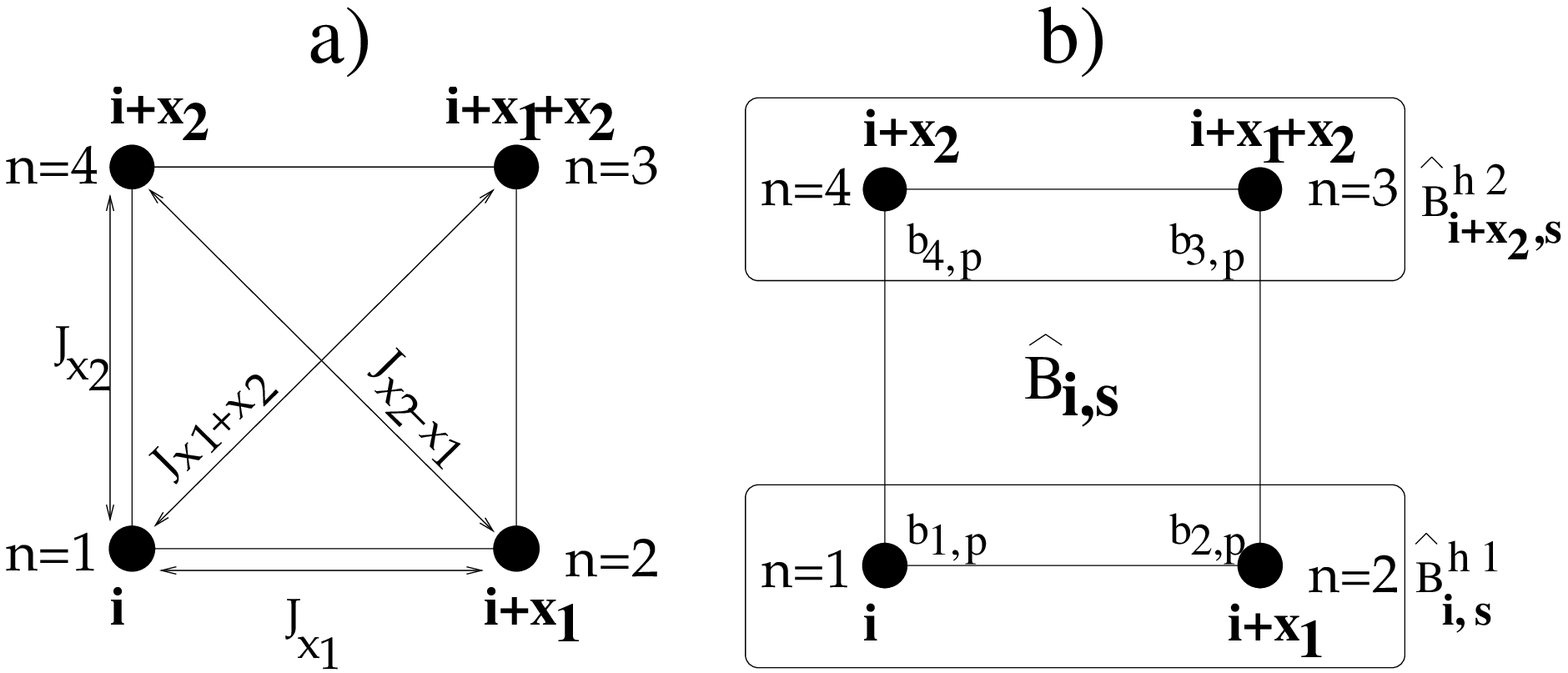}}
\caption{a)Unit cell ($\cal I$), and, b) the $\hat B_{{\bf i},s}$ operator
defined at an arbitrary site ${\bf i}$, ${\bf x}_{\protect\tau}$ are the
primitive vectors and $n$ is the ${\bf i}$ independent notation of the sites
in $\cal{I}_{\bf i}$. For a) arrows indicate hopping and hybridization matrix
elements (J=t, V). In b), the $b_{n,p}$, $p=d,f$
coefficients are shown together with a decomposition in horizontal
components $\hat B_{{\bf i},s} = \hat B^{h1}_{{\bf i},s} +
\hat B^{h2}_{{\bf i} + {\bf x}_2,s}$ of the complementary operator $\hat B_{
{\bf i},s}$.}
\label{fig1}
\end{figure}

Our starting Hamiltonian is thus,
$\hat{H}=\hat{H}_{0}+ \hat U$, written for a
free ($d$) and a correlated ($f$) band, where we allow for
hopping in both bands:
\begin{eqnarray}
&&\hat{H}_{0}=\sum_{{\bf i},\sigma }\biggm\{\sum_{{\bf r}}\left[
\sum_{p=d,f} t_{{\bf r}}^{p}\hat{p}_{{\bf i},\sigma }^{\dagger }
\hat{p}_{{\bf i}+{\bf r},\sigma}
+ V_{{\bf r}}(\hat{d}_{{\bf i},\sigma }^{\dagger }\hat{f}_{{\bf i}+%
{\bf r},\sigma } + \right.  \nonumber \\
&&\left. \hat{f}_{{\bf i},\sigma }^{\dagger }\hat{d}_{{\bf i}+{\bf r}%
,\sigma })+H.c.\right] +V_{0}(\hat{d}_{{\bf i},\sigma }^{\dagger }\hat{f}_{%
{\bf i},\sigma }+H.c.)+E_{f}\hat{n}_{{\bf i},\sigma }^{f}\biggm\}\: .
\label{Ham}
\end{eqnarray}
To keep the description as general as possible, the above Hamiltonian
is defined on a 2D Bravais lattice with unit cell
$\cal I$ and primitive vectors
$( {\bf x}_1$, ${\bf x}_2 )$. In Eq.\ (\ref{Ham})
$t_{{\bf r}}^{p}$, $V_{{\bf r}}$, $V_{0}$, and $E_{f}$, represent
the hopping amplitudes for $p=d,f$ electrons, the non-local and
on-site hybridization, and the local $f$ electron energy, respectively.
The coordinate ${\bf r}$, with possible values ${\bf x}_1, {\bf x}_2,
{\bf x_2} \pm {\bf x}_1$, is allowed to extend to {\em all} sites of $\cal I$.
Within the unit cell $\cal{I}_{{\bf i}}$ defined at ${\bf i}$,
the lattice sites
${\bf r}_{\cal{I}_{{\bf i}}}={\bf i}+{\bf r}_{\alpha \beta}$,
with ${\bf r}_{\alpha \beta } = \alpha {\bf x}_1 + \beta {\bf x}_2$,
$\alpha, \beta = 0,1$, can be numbered by $n(\alpha, \beta) = 1 + \alpha +
3 \beta - 2 \alpha \beta$ without reference to $\cal{I}_{\bf i}$,
see Fig.\ \ref{fig1} a).
The correlated $f$ band has in addition
$\hat U = U \sum_{{\bf i}}\hat{n}^{f}_{
{\bf i},\uparrow } \hat{n}^{f}_{{\bf i},\downarrow}$, acting on it with
$0< U < \infty$ on-site Hubbard repulsion.

It is known that exact solutions exist mostly in 1D systems and
in higher dimensions is almost impossible to find rigorous results.
Thus, to find the exact ground state of $\hat{H}$ in 2D we use a method
which is based on transforming $\hat{H}$ into a representation which is
positive semidefinite. Quantum mechanics tells us that the minimum
possible eigenvalue of a positive semidefinite operator, e.g., $\hat{O}$,
is zero. Hence the ground state, e.g., $\vert \Psi_g \rangle$ of $\hat{O}$
can be constructed from $\hat{O} \vert \Psi_g \rangle = 0$.
The fact that ground states containing intrinsic inhomogeneities
can be obtained in this manner for a ,,non-integrable''
\cite{noni} model as 2D-PAM does not come as a surprise,
since the applied procedure works even in disordered case \cite{noni1}.
We further note that the method is described in details in Ref.\
\onlinecite{pam1} and has been previously used to solve
generalized PAM type models at $3/4$ filling in 2D \cite{pam0} and even in
3D \cite{pam}. In the present case the transformation is performed
in 2D at a lower filling region, which is described in 2D for the first
time in this paper.

The transformation proceeds in the following way: first we transform
exactly the starting Hamiltonian in a positive semidefinite form. This
is accomplished with the use of the operators
\begin{eqnarray}
\hat{A}_{{\bf i},\sigma }^{\dagger }= \sum_{\alpha ,\beta =0,1}
\sum_{p=d,f}a_{n(\alpha ,\beta ),p}^{\ast }\hat{p}_{{\bf i}+
{\bf r}_{\alpha \beta },\sigma }^{\dagger },
\label{AA}
\end{eqnarray}
which are linear combination of the original fermionic operators acting on the
corners of an elementary plaquette , see Fig\ \ref{fig1} a). It can be
easily seen that
$\hat P =\sum_{{\bf i},\sigma }\hat{A}^{\dagger}_{{\bf i},\sigma }
\hat{A}_{{\bf i},\sigma }$ contains exactly the same operators from
Eq.\ (\ref{Ham}). Hence, properly choosing the coefficients $a_{n,p}$,
$n=n(\alpha,\beta)$, $\hat H_0$ from Eq.\ (\ref{Ham}) can
be written as  $\hat H_0 = \hat P + E_g$, where
$E_g$ is a constant. The proper mapping is
\begin{eqnarray}
\sum_{j=1}^{l_M} a_{m_j,p} a^{*}_{n_j,p'} = T^{p,p'}_{\nu,{\bf r}}(t^p_{\bf r},
t^{p'}_{\bf r}, V_{\bf r},E_f) \: ,
\label{cond}
\end{eqnarray}
where for the ${\bf r}=\alpha_1 {\bf x}_1 + \alpha_2 {\bf x}_2$ values
allowed by (1), and $p,p'=d,f$,  $T^{p,p'}_{\nu,{\bf r}}$ is given by
\begin{eqnarray}
&&T^{p,p'}_{\nu,{\bf r}}
= (1-\delta_{\nu,0})[\delta_{p,p'} t^p_{\bf r} +
(1-\delta_{p,p'}) V_{\bf r}] + \delta_{\nu,0}[\delta_{p,p'} (\delta_{p,d} K +
\delta_{p,f}(E_f+K)) + (1-\delta_{p,p'})V_0],
\nonumber\\
&&n_j=j \delta_{\nu,0} + (j+|\alpha_1-\alpha_2|)
\delta_{\nu,2} +
(\delta_{j,2} + (4|\alpha_1| + 2 |\alpha_2|)\delta_{j,1})
\delta_{\nu,1},
\nonumber\\
&&m_j-n_j = 2\delta_{\nu,2} + \delta_{\nu,1} [7-4j+(6j-10)
|\alpha_1| + (6j-8)|\alpha_2|].
\nonumber\\
&&2 l_M=8-5\nu+\nu^2, \quad \nu=(|\alpha_1|+|\alpha_2|).
\label{tpp}
\end{eqnarray}
Furthermore, $E_g=-K N$, where $N$ is the number of electrons,
and $K=\sum_{n=1}^4|a_{n,d}|^2$. Consequently, the starting Hamiltonian,
$\hat H$, becomes positive semidefinite
\begin{eqnarray}
\hat{H}= \hat P + \hat U + E_g \: ,
\label{HA}
\end{eqnarray}
except the additive constant $E_g$.
The transformation of the starting Hamiltonian based on the plaquette operators
from Eq.\ (\ref{AA}) into Eq.(\ref{HA}) is possible only if Eq.(\ref{cond})
containing 19 non-linear coupled equations allows solutions for the $a_{n,p}$
parameters. The different type of solutions of these non-linear equations
are presented in the following.

\section{Ground state solutions}

We found that there are two types of solutions
which satisfy the system of non-linear equations (\ref{cond}). These two types
of solutions will be denoted by $R = I$ and $R = II$. In
$I$, for $a^{*}_{n,d} /
a^{*}_{n,f}= q_n =q=$ real for all $n$, $\hat A_{{\bf i},\sigma}$
reduces to one-site form $\hat A_{{\bf i},\sigma} = \sum_{n=1}^4 a_{n,d}
\hat {\cal A}_{{\bf i}+{\bf r}_n,\sigma}$,   $\hat {\cal A}_{{\bf i},\sigma} =
\hat d_{{\bf i},\sigma} + \hat f_{{\bf i},\sigma}$ for all ${\bf i}$.
While in $II$,  $q_n \ne q =$real, and such a reduction of
$\hat A_{{\bf i},\sigma}$ into $\hat {\cal A}_{{\bf i},\sigma}$
is not possible.

In the following we will analyze in details the ground state
corresponding to the solutions $I$ and $II$:
(A) We first determine the homogeneous phases at
quarter filling ($N=N_{\Lambda}$, $N_{\Lambda}$ being the number of lattice
sites). (B) By decreasing $N$ we find degenerated droplet, stripe and
checkerboard ground states which we will present in detail,
and (C) extensions of $\hat H$ are identified to lift the
degeneracy leading to pure, non-degenerate
stripe and checkerboard ground states.

\subsection{Ground states at quarter filling}

To find the ground state at quarter filling,
a {\em complementary} unit cell
operator \cite{pam1} $\hat B_{{\bf i},\sigma}$ is defined by
\begin{eqnarray}
\{\hat A_{{\bf i},\sigma}, \hat B^{\dagger}_{{\bf i}',\sigma'} \} = 0 \: ,
\quad \forall \: {\bf i},{\bf i}',\sigma, \sigma' \: .
\label{ab}
\end{eqnarray}
For case $II$, $\hat{B}_{{\bf i},\sigma }=\sum_{\alpha ,\beta =0,1}
\sum_{p=d,f}b_{n,p} \hat{p}_{{\bf i}+{\bf r}_n,\sigma }^{\dagger }$,
[as shown in Fig.\ \ref{fig1} b)], and 
defining $n'(\alpha,\beta) = 3 + \alpha -\beta -2 \alpha
\beta$, and taking $w=b_{1,d}/a_{3,f}^{*}$, one finds
\begin{eqnarray}
b_{n,f} = - w a^{*}_{n',d}, \quad b_{n,d} = + w a^{*}_{n',f} \: .
\label{opb}
\end{eqnarray}
While, for case $I$,
$\hat B_{{\bf i},\sigma} = \sum_{n=1}^4 b_{n,d} \hat {\cal B}_{{\bf i} +
{\bf r}_n,\sigma}$, and $\{ \hat {\cal A}_{{\bf i},\sigma}, \hat {\cal B}^{
\dagger}_{{\bf i}',\sigma'} \} = 0$, $\hat {\cal B}_{{\bf i},\sigma} =
\hat d_{{\bf i},\sigma} - \hat f_{{\bf i},\sigma}$ holds.

Introducing $\hat D^{\dagger}_{{\bf i},\sigma} (R) = (
\hat {\cal C}^{\dagger}_{{\bf i},\sigma} \delta_{R,I} + \hat B^{\dagger}_{
{\bf i},\sigma} \delta_{R,II})$, where $\hat {\cal C}_{{\bf i},\sigma} = (
\hat {\cal B}^{\dagger}_{{\bf i},\sigma} + v_{\bf i} \hat {\cal B}^{\dagger}_{
{\bf i},-\sigma})$ and $v_{\bf i}$ are arbitrary coefficients, the exact
ground state at quarter filling becomes
\begin{eqnarray}
|\Psi_{g,R,1/4} \rangle = \prod_{{\bf i}=1}^{N_{\Lambda}}
\hat D^{\dagger}_{{\bf i},\sigma}(R) |0 \rangle \: ,
\quad R=I,II \:
\label{gr14}
\end{eqnarray}
where $|0\rangle$ is the bare vacuum.
This is the ground state wave function, i.e.,
$(\hat P + \hat U ) |\Psi_{g,R,1/4}\rangle = 0$. Here
$\hat U |\Psi_{g,R,1/4}\rangle = 0$
since, for $R=I$ the electrons with arbitrary spins are introduced on
different sites, and for $R=II$ all $\hat B^{\dagger}_{{\bf i},\sigma}$
in (\ref{gr14}) have the same fixed spin $\sigma$, so the double occupancy
for both $p$ is excluded. Besides, based on (\ref{ab}), further
we have $\hat P |\Psi_{g,R,1/4}\rangle = 0$. Since the minimum
eigenvalue of $\hat P + \hat U$ is zero, thus $|\Psi_{g,R,1/4}\rangle$ is
the ground state
for non-zero, positive, although arbitrary $U$. The ground state for
case $I$ is degenerate and globally paramagnetic, with one electron
on each site, i.e. the state is localized. While in case $II$ the ground
state is non-degenerate, is a
saturated ferromagnet, with $0$, $1$, or $2$ electrons on any
${\bf i}$ site, e.g. the state is itinerant.

\subsection{Ground states below quarter filling}

\subsubsection{Random droplet ground states}

\begin{figure}[h]
\centerline{\epsfbox{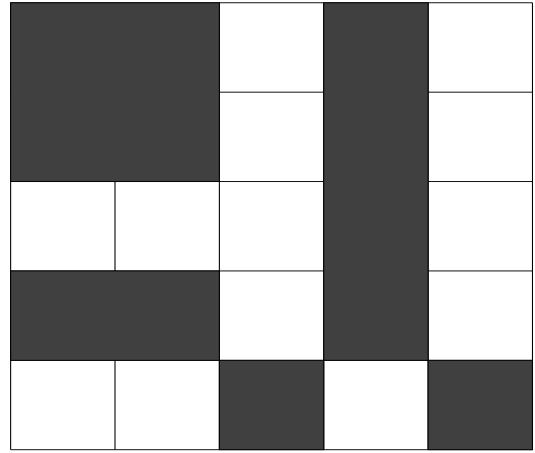}}
\caption{(a) Isolated plaquette with opposite spin at site $i=15$, and
(b) droplet solutions for case $II$. The blocks
$B_{i,\sigma}$ introducing the  $\hat B^{\dagger}_{{\bf i},\sigma}$ operators
in the ground state wave-function of case $II$ are presented as shaded
plaquettes in b).}
\label{fig2}
\end{figure}

Decreasing the number of $\hat D^{\dagger}_{{\bf i},\sigma}$ operators
in the product of Eq.(\ref{gr14}), the exact ground state below quarter
filling can be written as well. In case $I$, for $N < N_{\Lambda}$, we obtain
\begin{eqnarray}
|\Psi_{g,I, 1/4 \geq n > 0} \rangle =
\prod_{{\bf i}=1}^{N} \hat {\cal C}^{\dagger}_{{\bf i},\sigma} |0 \rangle \: ,
\label{e1}
\end{eqnarray}
where sites ${\bf i}$ can be arbitrarily chosen. Being empty sites present,
the ground state is an itinerant paramagnetic phase. In case $II$,
the ground state is reached only if touching
$\hat B^{\dagger}_{{\bf i},\sigma}$
operators defined on different sites [for example $(\hat B_{19,\sigma},
\hat B_{25,\sigma})$ in Fig.\ \ref{fig2} a)] have the same spin
in order to maintain
the $\hat U |\Psi_g\rangle = 0$ condition. These neighboring
$\hat B^{\dagger}_{{\bf i},\sigma}$ operators with fixed spin build up
different blocks (droplets) $Bl_j$ containing $N_{Bl_j}$ particles.
Two different blocks have no common lattice sites, and their spin
is non-correlated. If we denote by $N_{Bl}$ the number of blocks,
then the ground state wave function becomes
\begin{eqnarray}
|\Psi_{g,II,1/4 \geq n > 0} \rangle = \prod_{j=1}^{N_{Bl}}
\left( \prod_{{\bf i}=1}^{N_{Bl_j}} \hat B^{\dagger}_{
{\bf i},\sigma_j} \right) |0 \rangle \: ,
\label{e2}
\end{eqnarray}
where $\sum_j N_{Bl_j} = N$, but otherwise, $N_{Bl_j}$
and the shape of the blocks remain arbitrary, as on the example
shown in  Fig.\ \ref{fig2} b).
This state is itinerant and represents a fully saturated ferromagnet
up to $N \ge N_{\Lambda} - 8$.
In the limit
$N = N_{\Lambda} - 8$ there is always a possibility for a new block
to appear on the lattice, in a way that is not in contact with other blocks
and as such it can have opposite spin.
For example, in Fig.\ \ref{fig2} a), the middle block $B_{15,\downarrow}$
is isolated and has opposite spin compared to the big block surrounding it.
Hence, for $N < N_{\Lambda} - 8$ the ground state is not anymore
fully saturated. Further decreasing $N$, the ground state
remains ferromagnetic \cite{ferro} until two disjoint
blocks with the same number of sites but opposite spins can be constructed.
This happens at
$N_p^c = N_{\Lambda} - 2 L$, where $N_{\Lambda} = L \times L$.
Thus, for $N < N_p^c$, the ground state is globally paramagnetic.

\subsubsection{Degenerated stripe ground state solutions}

\begin{figure}[h]
\centerline{\epsfbox{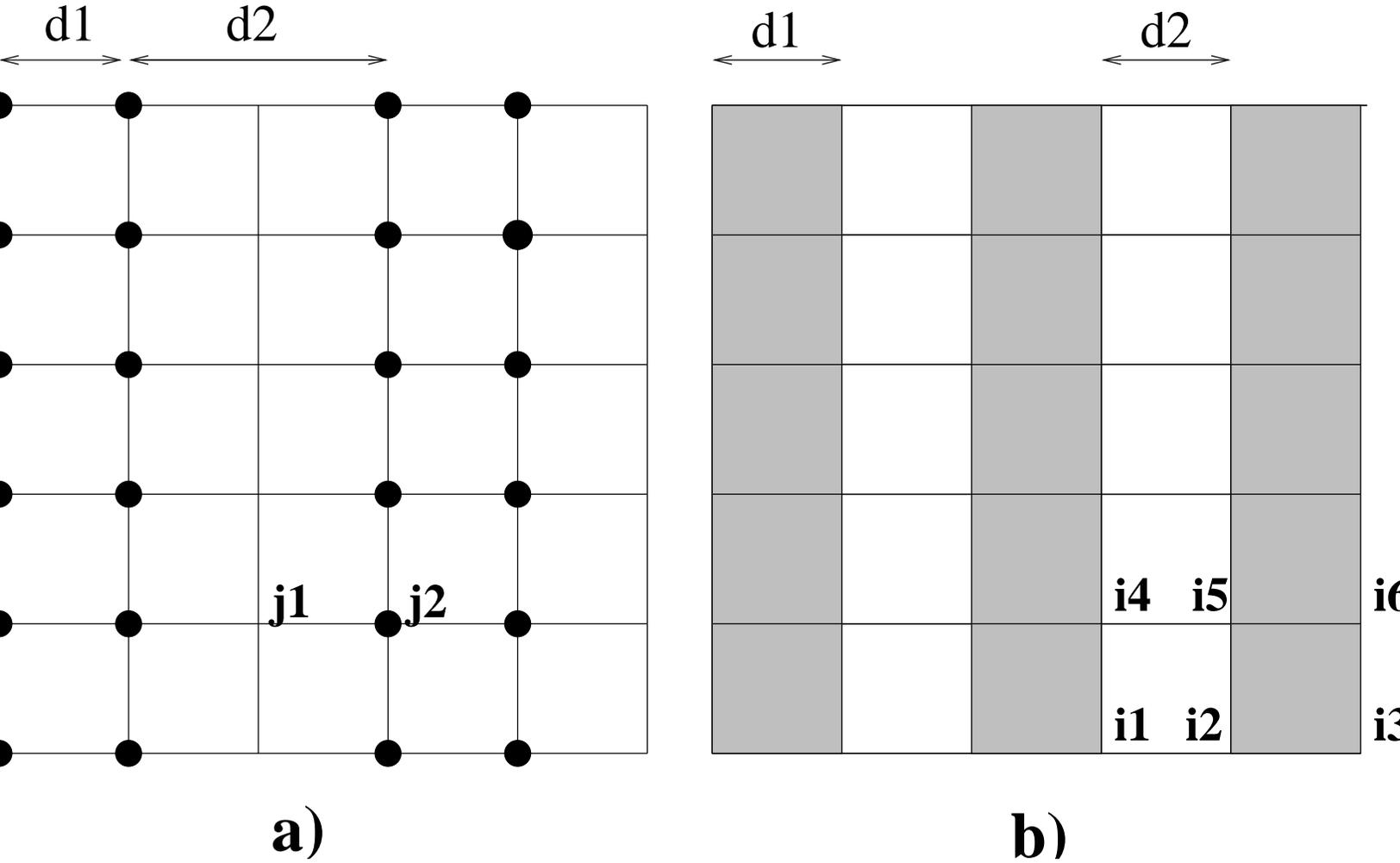}}
\caption{Example of vertical stripe solutions for case I shown in (a) and
II in (b).
For a) (or b)) full circles (or shaded plaquettes) denote sites
(or unit cells) whose
$\hat {\cal C}^{\dagger}_{{\bf i},\sigma}$  (or $\hat B^{\dagger}_{{\bf i},
\sigma}$) contribution is in the ground state wave-function Eq.(8).
For example, site $j2$ in case $I$ (unit cell (i2,i3,i5,i6) in case $II$).
In both cases $d1$ ($d2$) represents a measure of the stripe line width
(inter stripe line distance) in $x$ direction and $|{\bf x}_1|$ units.}
\label{fig3}
\end{figure}

\begin{figure}[h]
\centerline{\epsfbox{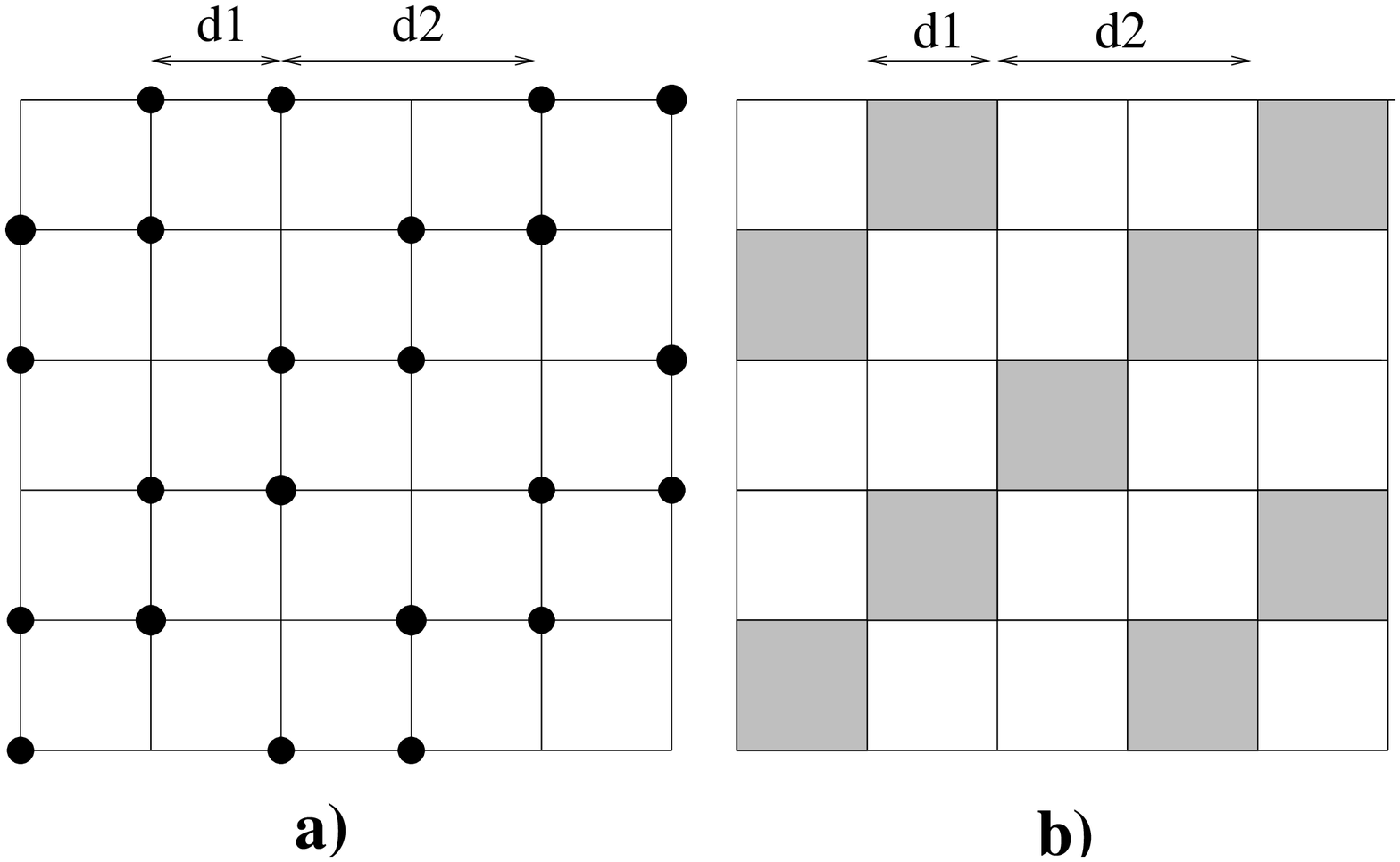}}
\caption{Example of diagonal stripe solutions for case $I$ shown in (a)
and case $II$ in (b).
Notations are as in Fig.3.}
\label{fig4}
\end{figure}

\begin{figure}[h]
\centerline{\epsfbox{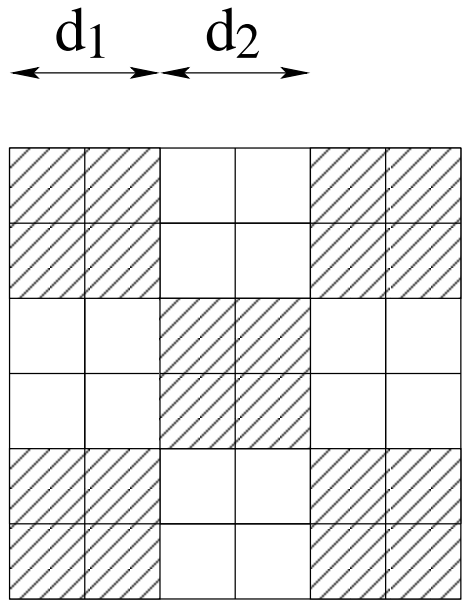}}
\caption{Checkerboard originating from diagonal stripes at $d_1=d_2$, see
also Fig.4.b)}
\label{fig5}
\end{figure}

Decreasing $N$ below $N_p^c$, stripes emerge in the ground state in
both Eqs.\ (\ref{e1},\ref{e2}) as vertical stripes (Fig.\ \ref{fig3})
or diagonal stripes (Fig.\ \ref{fig4}). The ground state wave function
containing $N_{St}$ vertical stripe lines is
\begin{eqnarray}
|\Psi_{g,R,I_{St}}\rangle = \prod_{j=1}^{N_{St}} ( \prod_{{\bf i} \in
I_{St,j}} \hat D^{\dagger}_{{\bf i},\sigma_j} ) |0\rangle \: ,
\label{e3}
\end{eqnarray}
where $I_{St,j}$ in case $I$ ($II$) represents the
stripe line $j$ (plaquette stripe column $j$), and $I_{St}= \sum_j I_{St,j}$.
For example, from (\ref{e3}), the vertical stripes are obtained in
case $I$ ($II$) by a displacement
along periodic vertical lines (vertically constructed plaquette
columns) of $\hat {\cal C}^{\dagger}_{{\bf i},\sigma}$  ($\hat B^{\dagger}_{
{\bf i},\sigma}$) operators. In Fig.\ \ref{fig3} a) case $I$ is shown
with column stripes, while Fig.\ \ref{fig3} b) depicts case $II$ with
plaquette columns.
For diagonal stripes, the vertical displacement
must be changed to diagonal one, as shown in Fig.\ \ref{fig4}. In the
case $II$ at $d1 = d2$, see Fig.\ \ref{fig5}, the stripe structure turns
into a checkerboard one. The stripe lines for case $I$ are
paramagnetic and insulating, while they are itinerant and ferromagnetic
for case $II$. Different stripe lines which are not in contact
have non-correlated spin.

\subsection{Non-degenerated stripe and checkerboard ground states}

For $N < N^c_p$, droplet [see Fig.\ \ref{fig2} b)] and stripe solutions
coexist as the ground state is degenerate. However, the droplet contributions
can be eliminated in exact terms from the ground state by adding new
Hamiltonian contributions to $\hat H$. For example,
let us consider $\hat H_A = - |W_1|\sum_{{\bf i} \in I_{St}}
\hat n_{\bf i}$, where $\hat n_{\bf i}=\sum_{\sigma,p} \hat n^p_{{\bf i},
\sigma}$. If $I'_{St}$ contains all lattice sites
from $\hat {\cal D}^{\dagger}_{I'_{St}} = \prod_{j=1}^N \hat D_{
{\bf i}_j,\sigma_j}$ and $I_{St}=I'_{St}$ holds, then
$[\sum_{{\bf i} \in I_{St}} \hat n_{\bf i}] \hat {\cal D}^{\dagger}_{I'_{St}}
= \hat {\cal D}^{\dagger}_{I'_{St}}[N + \sum_{{\bf i} \in I_{St}} \hat n_{
\bf i}]$ provides the minimum possible eigenvalue for $\hat H_A$ via
$\hat H_A |\Psi_{g,R,I_{St}}\rangle = - |W_1| N |\Psi_{g,R,I_{St}}\rangle$.
If however, $I'_{St} \ne I_{St}$ [for example, if
${\bf j}_2$ moves to the in ${\bf j}_1$ position, Fig.\ \ref{fig3} a)], then
$|\Psi_{g,R,I_{St}}\rangle$ is not anymore an eigenstate of $\hat H_A$.
Consequently, $|\Psi_{g,R,I_{St}}\rangle$ becomes the unique, non-degenerate
ground state of $\hat H + \hat H_A$. If we add to $\hat H_A$
the term $\hat H_A' =
|W_2|\sum_{{\bf i} \notin I_{St}} \hat n_{\bf i}$, as well, the
results remain unchanged. A Hamiltonian term of the form $\hat H_A$ is
motivated in case of cuprates by low temperature tetragonal
fluctuations \cite{mae}. The potential $W$ in
$\hat {\bar H}_A = \hat H_A + \hat H_A'$
can be generated by a periodic charge displacement or
charge density wave (see Fig.\ \ref{fig7}), which is
able to stabilize a stripe phase. Such a behavior has been already
seen \cite{cd} in cuprates. Similarly, the checkerboard state also
can be made stable \cite{ujcikk}. A checkerboard
is obtained, for example in Fig.\ \ref{fig4} b), if we take $d1 = d2 = 2$.
This is stabilized by a
Hamiltonian term $\hat H_A$, which has a set of lattice sites $I_{St}$, in
such a way that it contains only next-nearest-neighbor sites on each second
diagonal.

\begin{figure}[h]
\centerline{\epsfbox{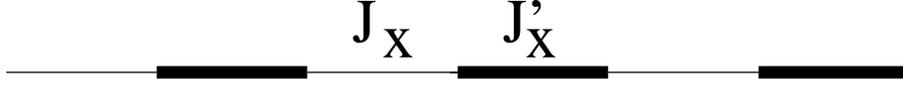}}
\caption{Bond alternation in x-direction to stabilize the stripes from
Fig. 3 b), $J_x=t_x^p,V_x$.}
\label{fig6}
\end{figure}

\begin{figure}[h]
\centerline{\epsfbox{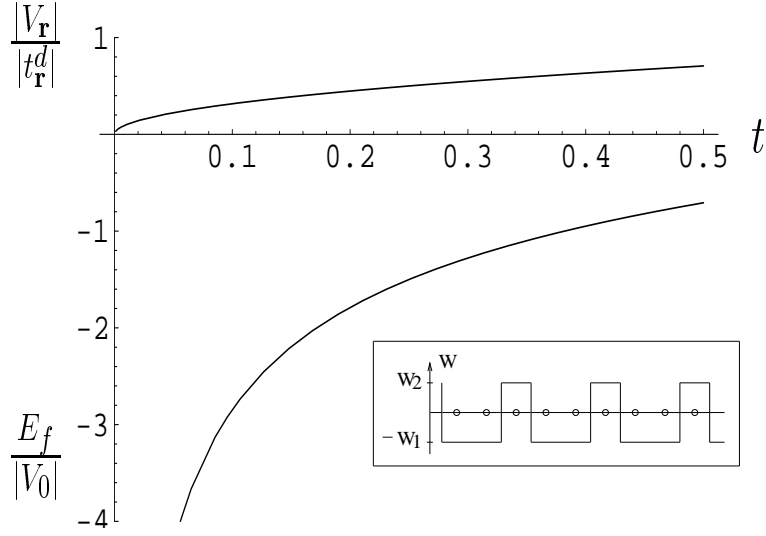}}
\caption{$\hat H_0$ parameters for case $I$ vs. $t=t^f_{{\bf x}_{\tau}}/t^d_{
{\bf x}_{\tau}}$, at $(t^d_x)^2, (t^d_y)^2 \geq
4 t^d_{x+y} t^d_{y-x}$. The inset shows the potential $W$
created by $\hat {\bar H}_A$ acting on the charge degrees of freedom in x
direction stabilizing the stripes from Fig.3 a).}
\label{fig7}
\end{figure}

\begin{figure}[h]
\centerline{\epsfbox{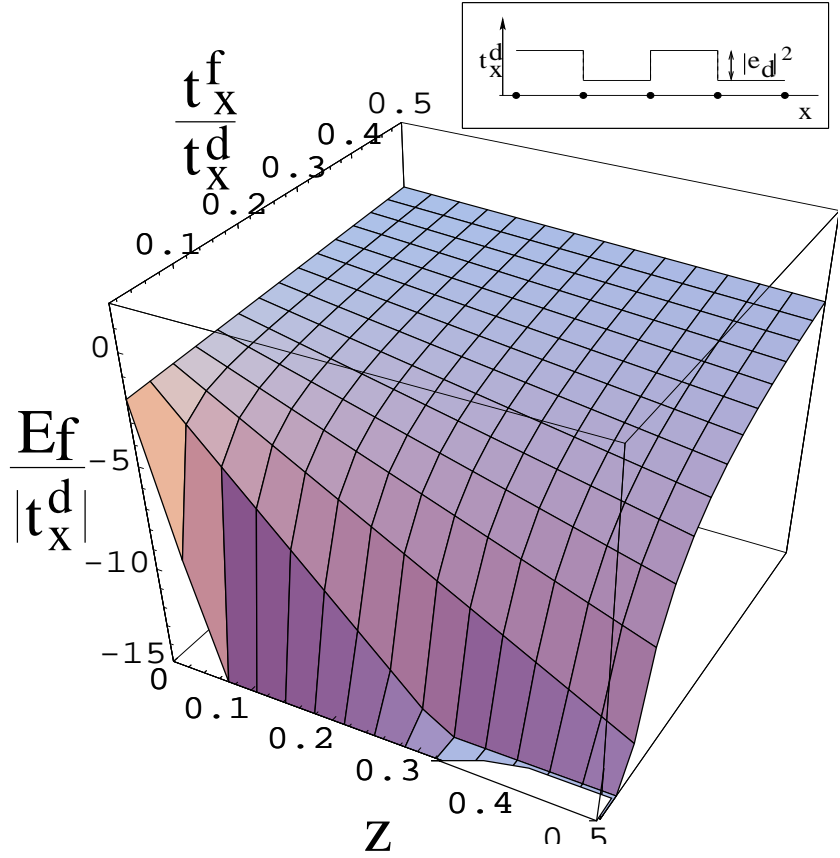}}
\caption{(Color online)
The $E_f/|t^d_x|$ surface in function of $t=t^f_x/t^d_x$ and the
anisotropy parameter $z=(|t^f_y|-|t^f_x|)/|t^d_x|$ above which the $R=II$
stripe ground-state from Fig.3.b emerge stabilized by $\hat H_B$
in conditions
from Ref.[23]. For hybridization $V_x=V_0=0$, and for other ${\bf r}$,
$|V_{\bf r}|/|t^f_{\bf r}|= \sqrt{|t|}$ holds. The inset shows the required
$t^d_x$ alternation introduced by $\hat H_B$. For $\hat H_B=0$, FM
is present on the plotted surface at $1/4$ filling.}
\label{fig8}
\end{figure}

Furthermore, the vertical stripes from Fig.\ \ref{fig3} b), will be
stabilized by a dimerization term of the form $\hat H_B = \sum_{\sigma}
\sum_{{\bf i} \in I_E} \hat E^{\dagger}_{{\bf i},\sigma} \hat E_{{\bf i},
\sigma}$, $\hat E_{{\bf i},\sigma} =
\sum_{p=d,f} (e_{1,p} \hat p_{{\bf i},\sigma} + e_{2,p} \hat p_{{\bf i} +
{\bf x}_1,\sigma})$, where $I_E$ contains only each second site of
horizontal lattice rows \cite{obst}, see Fig.\ \ref{fig6}.
For $\hat H_B$ described in Ref.\ \onlinecite{obst}, the stable
stripe phase occurs above the surface presented in Fig.\ \ref{fig8}.
Modifying $I_E$, other stripes can be addressed as well. The important
effect of dimerization on stripes stabilization have been recently shown
using numerical simulations \cite{nat}.

\section{Discussion of the results}

The obtained inhomogeneities are not connected in principle neither to 
certain special values of the parameters, nor to special properties of the
homogeneous phases present at quarter filling. In the following
we present several remarks to support this and in closing the
obtained stripe formation process is summarized.

Firstly, for a given Hamiltonian, a given decomposition in positive
semidefinite operators at a given filling factor is not unique
\cite{pam1,utx,utx1}. The decomposition itself can be done in
several different ways, leading to different $\hat A_{{\bf i},\sigma}$
operators, different matching conditions, hence different interdependencies
between microscopic parameters obtained during the solution of the matching
conditions (i.e. equations of the type (\ref{cond})), providing similar
solutions, and similar characteristics in different regions of the parameter
space. For example, as shown in Ref.[\cite{utx}], instead of plaquettes,
rhombi can also be used in defining $\hat A_{{\bf i},\sigma}$, obtaining the
same type of solutions, but in other regions of the parameter space. Other
possibilities are to simply tilt the unit cell \cite{pam1}, or to decompose in
$\sum_{\bf i} \hat A^{\dagger}_{\bf i} \hat A_{\bf i}$, where
$\hat A_{\bf i} = \sum_{\sigma} \hat A_{{\bf i},\sigma}$, as described in
Ref.[\cite{utx1}], etc.

Secondly, the different solutions of the matching conditions all give
homogeneous phase at quarter filling. Stripes will appear from each
of these homogeneous phases with (hole) doping. This can be easily understood
from the observation that the
$\hat B_{{\bf i},\sigma}$ (or $\hat {\cal B}_{{\bf i},\sigma}$) operators
which are characteristic of the stripe ground state, see eg,
Eqs.(\ref{ab},\ref{opb}), can appear for arbitrary $\hat A_{{\bf i},\sigma}$.
Consequently, the obtained stripe formation process
is weakly dependent on the properties of the homogeneous phase
from which the stripes emerge.

Thirdly, each homogenous phase (and hence each stripe structure originating
from it) appears for different, well-defined conditions.
For example, see also Ref.[\cite{ferro}],
stripes with ferromagnetic and itinerant stripe lines can be obtained from
the itinerant ferromagnetic homogenous phase when the one particle part of the
Hamiltonian in (\ref{Ham}) has such parameters that $\hat H_0$ will have a
diagonalized partially filled lower flat band. One the other hand, insulating
paramagnetic stripe lines are obtained in a different parameter space region,
where localized homogenous phase can occur, etc.

Fourthly, an additional repulsive Hubbard term acting in the $d$-band
will not change the obtained results. Consequently, the obtained
ground states and inhomogeneities are
valid not only in the 2D PAM, but also in a more general 2D two-band
Hubbard model as well.

Lastly, in our rigorous description we obtained a non-degenerate
ground state exhibiting stripe structure in the following steps:
we doped the homogeneous phase at quarter filling. This resulted
in a degenerate ground state to appear, which contained both
random droplets and stripe solutions. In the last step we
lifted the degeneracy by adding a so-called stabilization term
to the Hamiltonian. By doing this we eliminated the random droplets
from the degenerated ground state, obtaining providing a
non-degenerated stripe ground state solution.

\section{Conclusions}

In conclusion, providing exact results for
stripe, checkerboard vs. droplet interplay, we show how such
intrinsic inhomogeneities appear in a 2D Hamiltonian as a
non-degenerate ground states. For this, 1) a generalized PAM
is used and cast in a positive semidefinite form, 2) the ground states
are explicitly constructed at and below quarter filling, and
3) the ground state degeneracy provided by
random droplets is eliminated using extension terms representing, e.g.,
distortion lines, dimerization or density waves.
The inhomogeneities were shown to exist
in a broad concentration range below quarter filling and they are
either (case $I$) paramagnetic and localized,
or (case $II$) itinerant and ferromagnetic. In both of these cases the
inter-stripe lines are insulating. As has been underlined, stripes can
emerge from different homogeneous phases so are less sensitive to microscopic
parameters of $\hat H$. As marginal for the stripes, but important for PAM we
derive an exact ferromagnetic ground state, as well, in $2D$ at quarter
filling. An extra Hubbard interaction in the $d$ band,
$U_d > 0$ leaves the above results unchanged.
The presented positive semidefinite decomposition
is not unique, can be effectuated in several ways,
leading to similar results also in other regions of the parameter space.

\acknowledgements

We acknowledge valuable discussions with D. Vollhardt and J.
Zaanen. For Zs.G. research supported by grants OTKA-T037212 and K60066 of
Hungarian Scientific Research Fund, and Alexander von Humboldt foundation 
at Institute for Theoretical Physics III, University of 
Augsburg.


\newpage

\newpage

\newpage

\newpage

\newpage

\newpage

\newpage

\begin{thebibliography}{99}
\bibitem{noni}
In special cases called ,,integrable'' (i.e. the number of degrees of freedom
is equal to the number of constants of motion), the whole spectrum can be
deduced using methods as Bethe Ansatz. In our case, only the ground state is
obtained.
\bibitem{ferro}
Eq.(\ref{cond}) shows that when ferromagnetism emerges,
the diagonalized $\hat H_0$ has a partially filled lower flat band.
\end{thebibliography}
\end{document}